\documentclass[aps,prl,twocolumn,superscriptaddress]{revtex4}

\usepackage{mathrsfs}
\usepackage{graphicx}
\usepackage{bm}
\usepackage{threeparttable}
\usepackage{amsmath,amssymb}
\usepackage{color}
\usepackage{epstopdf}
\usepackage[hidelinks]{hyperref}

\begin{document}

\title{Surface Tension Between Coexisting Phases of Active Brownian Particles}

\author{Longfei Li}
\email{longfeili@iphy.ac.cn}
\affiliation{Beijing National Laboratory for Condensed Matter Physics and Laboratory of Soft Matter Physics, Institute of Physics, Chinese Academy of Sciences, Beijing 100190, China}

\author{Zihao Sun}
\affiliation{Beijing National Laboratory for Condensed Matter Physics and Laboratory of Soft Matter Physics, Institute of Physics, Chinese Academy of Sciences, Beijing 100190, China}
\affiliation{School of Physical Sciences, University of Chinese Academy of Sciences, Beijing 100049, China}

\author{Mingcheng Yang}
\email{mcyang@iphy.ac.cn}
\affiliation{Beijing National Laboratory for Condensed Matter Physics and Laboratory of Soft Matter Physics, Institute of Physics, Chinese Academy of Sciences, Beijing 100190, China}
\affiliation{School of Physical Sciences, University of Chinese Academy of Sciences, Beijing 100049, China}

\begin{abstract}
The confliction between the stable interface in phase-separated active Brownian particles and its negative surface tension, obtained mechanically via the active pressure, has sparked considerable debate about the formula of active surface tension. Based on the intrinsic pressure of active system, we here derive a new mechanical expression of active surface tension by calculating the work required to create a differential interface area, while remaining the interfacial profiles of intensive quantities invariant (not considered previously). Our expression yields a significantly positive surface tension that increases with the particle activity, which is further supported by mechanical stability analysis of both steady-state droplet and fluctuating interface. Our work is thus promising to resolve the contradiction related to active surface tension.
\end{abstract}

\pacs {}
\maketitle

\section*{Introduction}
Active matter consisting of an ensemble of self-propelled units is an important class of non-equilibrium systems~\cite{Ramaswamy2010arcmp,Marchetti2013rmp,Bechinger2016rmp} and can exhibit diverse exotic phenomena, forbidden in thermal equilibrium~\cite{Vicsek1995prl,Narayan2007sci,Cates2015arcmp,Dombrowski2004prl,Solon2015np,Di2010pnas,Liu2020prl,Yang2021prl}. A prominent example is motility-induced phase separation (MIPS), which refers to a spontaneous gas-liquid phase separation of purely repulsive active Brownian particles (ABPs) above certain densities and activities~\cite{Tailleur2008prl,Fily2012prl,Redner2013prl,Buttinoni2013prl,Cates2015arcmp}. Although great progress in the MIPS has been made in the past decade, surface tension of the active gas-liquid interface, a central physical quantity of the MIPS, still remains elusive and controversial.

Since Bialk\'e \textit{et al.} reported that the surface tension $\gamma$ is significantly negative in the phase-separated repulsive ABPs~\cite{Bialke2015prl}, a proper identification of the active $\gamma$ and the corresponding mechanism of interfacial stability have been widely debated~\cite{Speck2016epl,Patch2018sm,Das2020sm,Omar2020pre,Lauersdorf2021sm,Paliwal2017jcp,Chacon2022sm,Speck2020sm,Hermann2019prl,Speck2021pre,Wittmann2017jsmte,Solon2018njp,Tjhung2018prx,Fausti2021prl,Besse2023prl,Cates2023prl,Zhao2024arxiv}. In the work by Bialk\'e~\textit{et al.}, $\gamma$ is derived mechanically based on \textit{active pressure} and the resulting expression is the integration of the difference between normal and tangential active pressures across the surface region, formally the same as the Kirkwood-Buff equation~\cite{Kirkwood1949jcp}. Negative surface tension was repeatedly reported by subsequent studies~\cite{Speck2016epl,Patch2018sm,Das2020sm}. Intuitively, a negative $\gamma$ destabilizes the interface and thus contradicts with the stable MIPS. 

The mechanical determination of active $\gamma$ needs to identify the pressure of active systems, and however a consensual interpretation of the latter has yet been reached. Recently, it has been gradually recognized that the \textit{intrinsic pressure} in an active system should be composed of the ideal-gas pressure and the one arising from interparticle interactions (i.e., corresponding to the traditional pressure definition)~\cite{Yan2015sm,Speck2016pre,Steffenoni2017pre,Paliwal2018njp,Epstein2019jcp,Omar2020pre,Lauersdorf2021sm,Omar2023pnas}; while the active pressure aforementioned is the sum of the intrinsic pressure and the swim pressure~\cite{Takatori2014prl,Solon2015np,Solon2015prl,Winkler2015sm}. Within the framework of the intrinsic pressure, the coexisting gas and liquid phases have different pressures, that are balanced by the polarization-induced body force at the interface. With the intrinsic pressure, Omar \textit{et al.} derived a different mechanical expression of active surface tension, yielding a near-zero but still negative $\gamma$~\cite{Omar2020pre}. Moreover, combining the intrinsic pressure and Laplace-like equation, Lauersdorf \textit{et al.} also obtained a small $\gamma$ but with considerable uncertainty in the sign~\cite{Lauersdorf2021sm}. 

On the other hand, despite the out-of-equilibrium nature of active matter, capillary wave theory (CWT), one of the standard way to determine $\gamma$ in passive systems, is employed to quantify the active surface tension~\cite{Bialke2015prl,Speck2016epl,Patch2018sm,Paliwal2017jcp,Langford2024pre}. Therein, the CWT assumes an equilibrium-like equipartition relation, in which the traditional thermal energy is replaced by ``housekeeping work" or ``effective thermal energy''. Such an assumption, however, could be invalid in active systems. Another type of approach to calculate the active surface tension, following the equilibrium thermodynamic route, is to construct nonequilibrium chemical potential or effective free energy for the phase-separated ABPs~\cite{Hermann2019prl,Wittmann2017jsmte,Speck2021pre,Solon2018njp}. Some of them give a positive effective $\gamma$, while other may support both positive and negative values~\cite{Solon2018njp}. Nevertheless, the equilibrium-like effective thermodynamic treatments are elusive and not \textit{a priori} guaranteed by basic thermodynamic principles, such that the resulting effective $\gamma$ is generally not equivalent to its mechanical counterpart that is well-defined even far from equilibrium. In addition, scalar field theory has been used to investigate the surface tension for generic continuum active model in a coarse-grained way~\cite{Tjhung2018prx,Fausti2021prl,Besse2023prl,Cates2023prl}. It also allows both positive and negative effective $\gamma$, which however has no direct connection to the mechanical $\gamma$. Therefore, up to now, it is still a challenging open problem to unambiguously determine the %intuitively consistent 
mechanical surface tension of the phase-separated ABPs. 

In this article, we derive a new formula for the mechanical surface tension of the phase-separated ABPs by using the intrinsic pressure to calculate the work required to create a differential interface area. In our derivation, it is crucial to maintain the interfacial profiles of intensive physical quantities when changing the interfacial area, which is a natural requirement to examine the interface with exactly specified properties but is omitted in previous studies. Based on the obtained formula, our simulation results show that the active $\gamma$ is significantly positive, consistent with the stable interface observed in the MIPS. Furthermore, our findings are robustly corroborated through direct analyses of the mechanical stability of steady-state droplet and fluctuating interfaces.

\section*{Results}
\subsection*{A. Theory}
\begin{figure}[t]
	\includegraphics[width=0.49\textwidth]{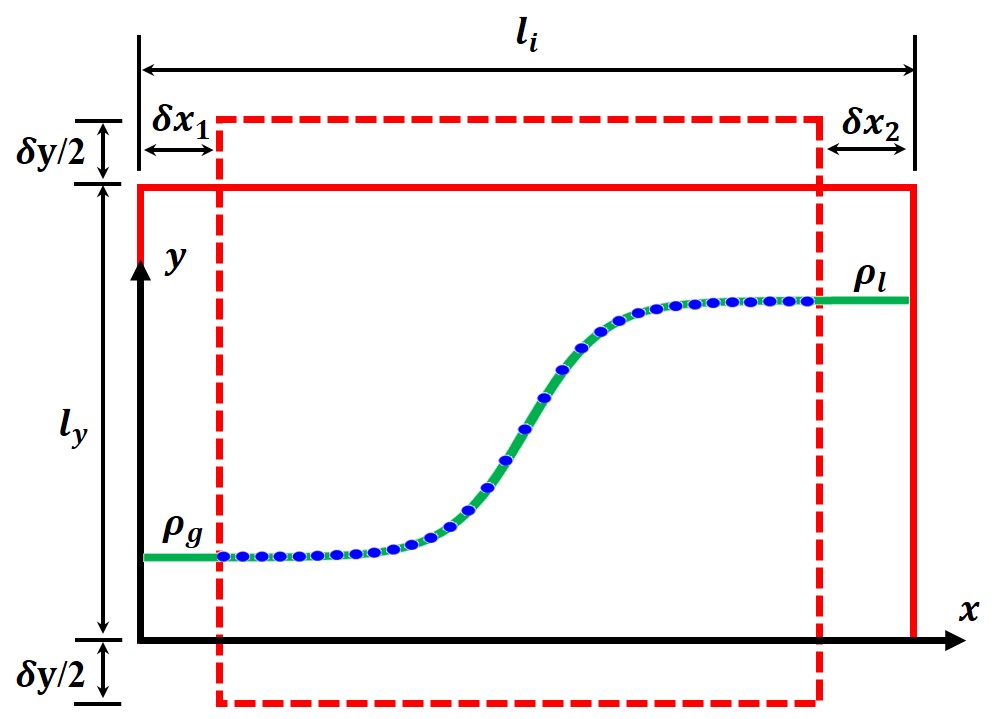}
	\caption{Schematic of the deformation of the surface region used to derive the surface tension. The red solid and dashed squares separately represent the original and deformed surface regions, with the $x$ axis perpendicular to the surface. The green solid and blue dotted curves denote the density profiles $\rho_i(x)$ of ABPs in the original and deformed surface regions, respectively, and they coincide with each other when the deformed surface region lacks a net whole movement.}
	\label{Fig1}
\end{figure}

We first derive the mechanical expression of the surface tension of the phase-separated ABPs. For convenience, the following calculation is confined to two dimensions, which can be straightforwardly extended to three-dimensional systems. The coexisting system consists of a homogeneous gas phase, a homogeneous liquid phase and an inhomogeneous surface region sandwiched between two bulk phases. The bulk pressure and density are separately $P_g$ and $\rho_g$ for the gas phase, and $P_l$ and $\rho_l$ for the liquid phase. Note that the pressure in the present work represents the \textit{intrinsic pressure}, unless stated otherwise. To analyze the surface properties, we isolate the surface region from the whole system, as sketched in Fig.~\ref{Fig1} by the red solid square of length $l_i$ and width $l_y$, with the $x$ axis perpendicular to the interface. The boundaries of the surface region are taken far into each bulk phase. As a result, the density profile $\rho_i(x)$ of the surface region (represented by the green solid curve in Fig.~\ref{Fig1}) saturates to the corresponding bulk-phase values at the positions far from the left and right borders of the surface region (i.e., $x=0$ and $x=l_i$), as well as the pressure profile $P_i(x)$ (not shown).

An important difference from a passive interface is that the active interface region possesses a polarization-induced body force density, $f_0 m(x)\hat{x}$, pointing to the dense phase \cite{Omar2020pre,Lauersdorf2021sm,Omar2023pnas}. Here, $f_0$ and $m(x)\hat{x}$ are the self-propelling force and the local polarization of active particles, respectively. The latter is given by $m(x) = (1/l_y)\int P(x, \theta) \cos {\theta}{\rm d} \theta$, where $P(x, \theta)$ denotes the probability distribution function of finding a particle locating at $x$ with an orientation $\theta$ (with respect to the $x$ axis). In the steady state, the polarization force on the surface per unit length, $f_0 M=\int_0^{l_i} f_0 m(x) {\rm d}x$, is balanced by the intrinsic pressure difference between the coexisting gas and liquid, i.e., $P_l - P_g = f_0 M$.

To obtain $\gamma$ from its mechanical definition, that is the work required to create a unit area of the surface with the volume fixed, we isothermally and quasistatically deform the surface region. We expand the surface region along the $y$ direction by $\delta y$ ($\delta y \ll l_y$), and meanwhile compress it along the $x$ direction by $\delta x_1$ and $\delta x_2$ separately from its left and right boundaries, as displayed by the red dashed square in Fig.~\ref{Fig1}. To maintain a constant volume, we have
\begin{equation}
	\label {eq1}
	\begin{split}
		l_i l_y = (l_i - \delta x_1 - \delta x_2)(l_y + \delta y).
	\end{split}
\end{equation}
For an equilibrium phase-separated state ($P_g=P_l$), the derivation of $\gamma$ does not depend on the specific values of $\delta x_1$ and $\delta x_2$, as long as $\delta x_1+\delta x_2$ satisfies Eq.~(\ref{eq1}) [as can be seen from Eq.~(\ref{eq3})]. However, for the MIPS ($P_g\neq P_l$), $\delta x_1$ and $\delta x_2$ need to be specified through an extra condition in order to perform a proper derivation. The extra important condition, that is usually neglected but needs to be met during the deformation of the system, is that the profiles of all intensive quantities (including the density profile) in the surface region must remain perfectly the same as their counterparts of the original interface. Otherwise, the system intensive quantities and hence the system properties will change (namely the interface we are examining is no longer the original one), thus giving rise to an improper $\gamma$. In other words, only under this condition can we obtain the true surface tension of the original system. Thus, the combination of the invariant density profile and conservation of particles yields
\begin{equation}
	\label {eq2}
	\begin{split}
		l_y \int_0^{l_i} \! \rho_i(x) \, {\rm d}x  = (l_y + \delta y) \int_{\delta x_1}^{l_i- \delta x_2} \! \rho_i(x) \, {\rm d}x.
	\end{split}
\end{equation}
Here, for simplicity, we have additionally prescribed that the deformed surface region experiences no net whole displacement with respect to the original surface region (as indicated by their overlapping density profiles in Fig.~\ref{Fig1}), so that the polarization force does no work.

For the above deformation, the work $\delta W$ done on the surface region can be easily written as
\begin{equation}
	\label {eq3}
	\begin{split}
		\delta W = P_g l_y \delta x_1 + P_l l_y \delta x_2 - \delta y\int_0^{l_i} P_i^T(x) \, {\rm d}x,
	\end{split}
\end{equation}
with $P_i^T(x)$ the tangential pressure (in the $y$ direction). From Eqs.~(\ref{eq1}-\ref{eq3}) and the mechanical definition, $\gamma = \delta W / \delta y$, neglecting higher-order terms, we obtain
\begin{equation}
	\label {eq4}
	\begin{split}
		\gamma = \int_0^{l_i} \! \left[ P_g\frac{\rho_l - \rho_i(x)}{\rho_l - \rho_g} + P_l \frac{\rho_i(x) - \rho_g}{\rho_l - \rho_g} - P_i^T(x) \right] {\rm d}x.
	\end{split}
\end{equation}
Equation~(\ref{eq4}) is the central result of this article, and it is valid in both equilibrium and nonequilibrium systems. It should be pointed out that for a more general deformation of the surface region that exhibits a net whole movement (say $\delta x'$), the pressures and the polarization-induced force seemingly do extra work on the interface region, $(P_g + f_0 M - P_l)l_y\delta x'$, besides the $\delta W$ of Eq.~(\ref{eq3}). Nevertheless, since $P_l = P_g + f_0 M$, the resulting surface tension is exactly the same as Eq.~(\ref{eq4}). 

It would be instructive to compare the present result, Eq.~(\ref{eq4}), with other existing theories. In equilibrium state, $P_g = P_l = P$, Eq.~(\ref{eq4}) reduces to the well-known Kirkwood-Buff expression~\cite{Kirkwood1949jcp},
\begin{equation}
	\label {eq5}
	\begin{split}
		\gamma = \int_0^{l_i} \! \left[ P - P_i^T(x) \right] {\rm d}x.
	\end{split}
\end{equation}
On the other hand, in the ABPs system, by only considering the area conservation of surface region in Eq.~(\ref{eq1}) but omitting the condition in Eq.~(\ref{eq2}), Omar \textit{et al.} recently obtained a similar formula to Eq.~(\ref{eq5}), but with the bulk-phase $P$ replaced by the local normal pressure $P(x)$~\cite{Omar2020pre}. And, their formula yields a small negative $\gamma$. Furthermore, by using the active pressure $P^A$ (in this case $P_g^A = P_l^A = P^A$), Bialk\'e \textit{et al.} reported a surface tension with the same form as Eq.~(\ref{eq5})~\cite{Bialke2015prl}, nevertheless their $\gamma$ is significantly negative. These negative $\gamma$ contradict with our physical intuition that a mechanically stable active surface, extensively observed in the MIPS, should possess a positive surface tension.

It is not convenient to directly use Eq.~(\ref{eq4}) in simulation to compute the surface tension, since it needs to track the surface location frequently. A numerically more efficient method is achieved by extending the interval of integration in Eq.~(\ref{eq4}) to the whole system (including the bulk phases and surface region), because of the null contribution outside the surface region, which results in
\begin{equation}
	\label {eq6}
	\begin{split}
		\gamma = \frac{1}{2} l_x \left[ \frac{P_g\rho_l - P_l \rho_g}{\rho_l -\rho_g} + \frac{P_l - P_g}{\rho_l -\rho_g} \rho_0 - \bar{P}^T  \right].
	\end{split}
\end{equation}
Here, the prefactor $\frac{1}{2}$ accounts for two surfaces due to periodic boundary condition, $l_x$ refers to the total length of the whole system in the $x$ direction, and $\rho_0$ and $\bar{P}^T$ are, respectively, the mean density and the average tangential pressure of the whole system.

\subsection{B. Simulations}

\begin{figure}[t]
	\includegraphics[width=0.47\textwidth]{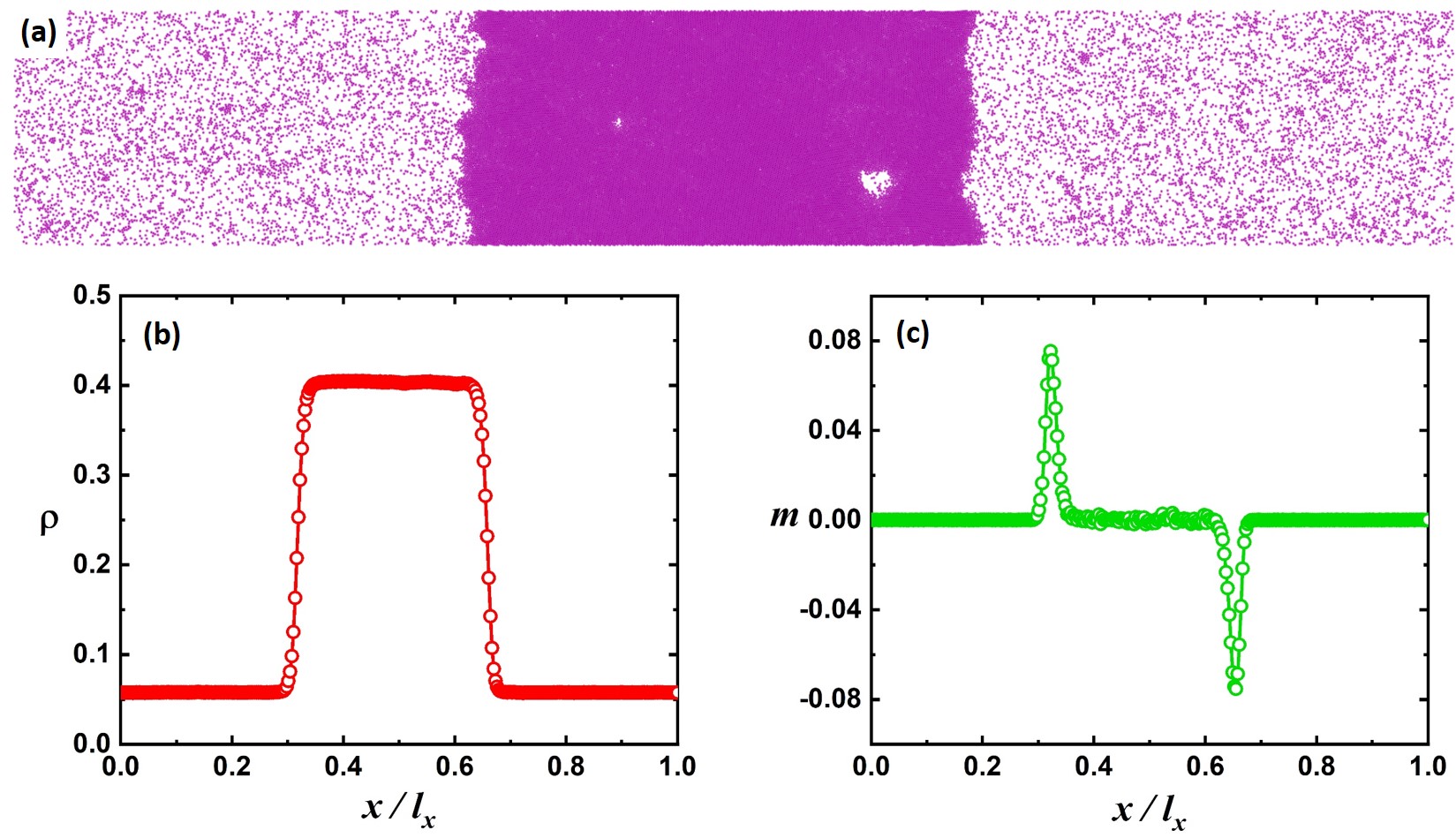}
	\caption{MIPS of the ABPs at $\text{Pe}=160$ and the mean density $\rho_0=0.175$. (a) A typical snapshot, (b) the time-averaged density profile and (c) polarization profile in a single run.}
	\label{Fig2}
\end{figure}

\textbf{B.1 Planar interface}. We now implement simulations to compute the surface tension of the phase-separated ABPs through Eq.~(\ref{eq6}). The MIPS is formed by $N=50000$ ABPs of diameter $\sigma$ in a rectangular box of size $l_x \times l_y$, with periodic boundary conditions. The interparticle interactions are described by the Weeks-Chandler-Andersen potential, $U(r)=4\epsilon[\left(\sigma/r\right)^{12}-\left(\sigma/r\right)^{6}]+\epsilon$ if $r<2^{1/6}\sigma$, and $U(r)=0$ otherwise. The motion of particle $i$ with orientation $\textbf{n}_i=[\cos \theta_i, \sin \theta_i]$ follows the overdamped Langevin equation,
\begin{equation}
	\label {eq7}
	\begin{split}
		\dot{\textbf{r}}_i &= v_0\textbf{n}_i - \frac{1}{\gamma_t}\nabla_{\textbf{r}_i}U_t + \sqrt{2D_t}\bm{\xi}_i; \quad
		\dot{\theta}_i = \sqrt{2D_r}{\zeta}_i.
	\end{split}
\end{equation}
Here, $D_t = k_BT/\gamma_t$ and $D_r = k_BT/\gamma_r$ separately are translational and rotational diffusion coefficients with the translational (rotational) friction coefficient $\gamma_t$ ($\gamma_r=\sigma^2\gamma_t/3$) and thermal energy $k_BT=\epsilon$, $U_t=\sum_{j} U(r_{ij})$, and $v_0=f_0/\gamma_t$ is a constant self-propelling velocity. In Eq.~(\ref{eq7}), $\bm{\xi}$ and $\zeta$ refer to the Gaussian-distributed white noises of zero mean and unit variance. In simulations, we take $\sigma=2$, $k_B T=1$ and $\gamma_t=100$, and other physical quantities are accordingly reduced. Particularly, the surface tensions $\gamma$ are reduced by $2k_B T/\sigma$, unless stated otherwise. We parameterize our system with P\'eclet number $\text{Pe}=3v_0/(\sigma D_r)$ and the mean density $\rho_0=N/(l_xl_y)$, and set $l_x/l_y \approx 6$ to encourage the coexisting dense phase to span the shorter dimension. To reach the steady state quickly, the particles are initially arranged in a slab with a hexagonal close-packed structure.

To evolve the system, Eq.~(\ref{eq7}) is integrated using the Euler scheme with a time step $dt=2\times 10^{-3}$, and each run lasts $5\times10^8$ time steps at least (after a relaxation of $5\times10^7$ time steps, ensuring that the system reaches the steady state). Note that the intrinsic pressure used to calculate the $\gamma$ consists of the ideal-gas pressure and the one due to interparticle interactions, the latter of which can be computed from the cross-correlation of the connecting vector between two particles and the corresponding pair force~\cite{Bialke2015prl}.

\begin{figure}[t]
	\includegraphics[width=0.43\textwidth]{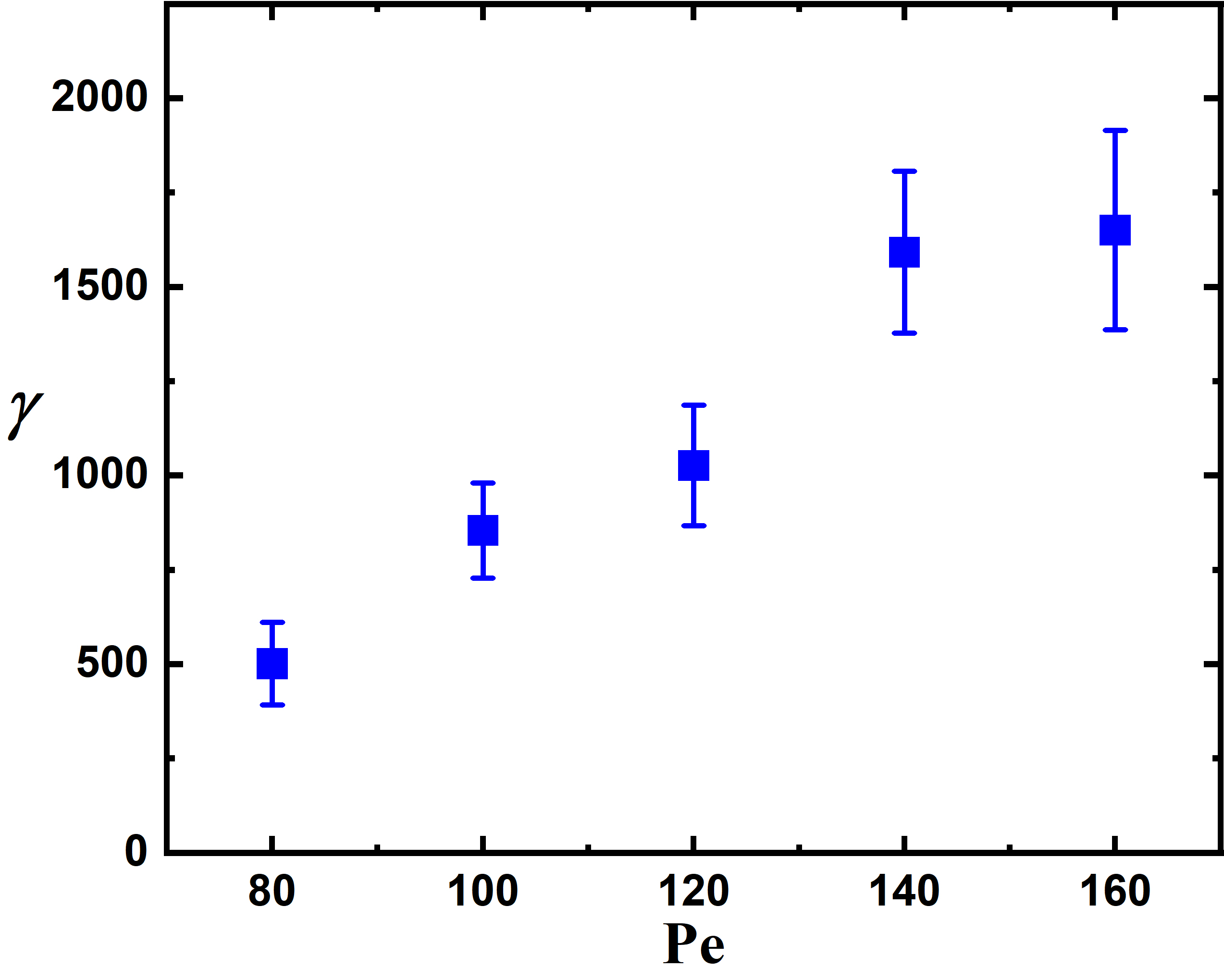}
	\caption{Active surface tension $\gamma$ in the MIPS as a function of $\text{Pe}$, obtained from the simulation based on Eq.~(\ref{eq6}). The data points with low and medium activities (from $\text{Pe}=80$ to $\text{Pe}=120$) are the averages over 10 independent measurements, while the results for high activity ($\text{Pe}=140$ and $\text{Pe}=160$) correspond to the averages of 6 independent simulations. Error bars denote the standard deviations.}
	\label{Fig3}
\end{figure}

Figure~\ref{Fig2}(a) shows a snapshot of the gas-liquid phase separation, and Figs.~\ref{Fig2}(b) and (c) are the corresponding profiles of density and local polarization along the $x$ axis, respectively. The density difference between the bulk gas and liquid of the ABPs means a difference in the intrinsic pressure, which is balanced by $f_0 M$, as demonstrated in the Supplementary material~\cite{suppl}. Using Eq.~(\ref{eq6}), we compute the active surface tension for varying $\text{Pe}$ realized by tuning the self-propelling velocity at a fixed mean density $\rho_0=0.175$. The result in Fig.~\ref{Fig3} shows that the $\gamma$ is much larger than zero and increases with $\text{Pe}$. This significantly positive $\gamma$ is probably caused by the great polarization-induced force occurring in the surface region. A positive $\gamma$ is consistent with the stable MIPS of the ABPs, as expected intuitively. Thus, we recover the traditional knowledge that a stable surface even in active systems should have a positive mechanical surface tension. 

To better compare with the surface tension previously quantified via the active pressure, we also simulate the MIPS system having the same parameters as those employed by Bialk\'e \textit{et al.}~\cite{Bialke2015prl}. In this case, the computed $\gamma$ from Eq.~(\ref{eq6}) is still significantly positive, $\gamma \approx 1067$, in striking contrast to the largely negative surface tension, $\gamma \approx -475$, obtained by Bialk\'e \textit{et al.}~\cite{Bialke2015prl}. Furthermore, we separately employ the schemes of Bialk\'e \textit{et al.}~\cite{Bialke2015prl} and Omar \textit{et al.}~\cite{Omar2020pre} to calculate $\gamma$ for our system with $\text{Pe}=120$. The former still yields a significantly negative surface tension, $\gamma \approx-883$; while, as expected, the latter leads to a near-zero surface tension, $\gamma \approx 0.016$, which is nondimensionalized in the same way as done in Reference~\cite{Omar2020pre}. 

\begin{figure}[t]
	\includegraphics[width=0.47\textwidth]{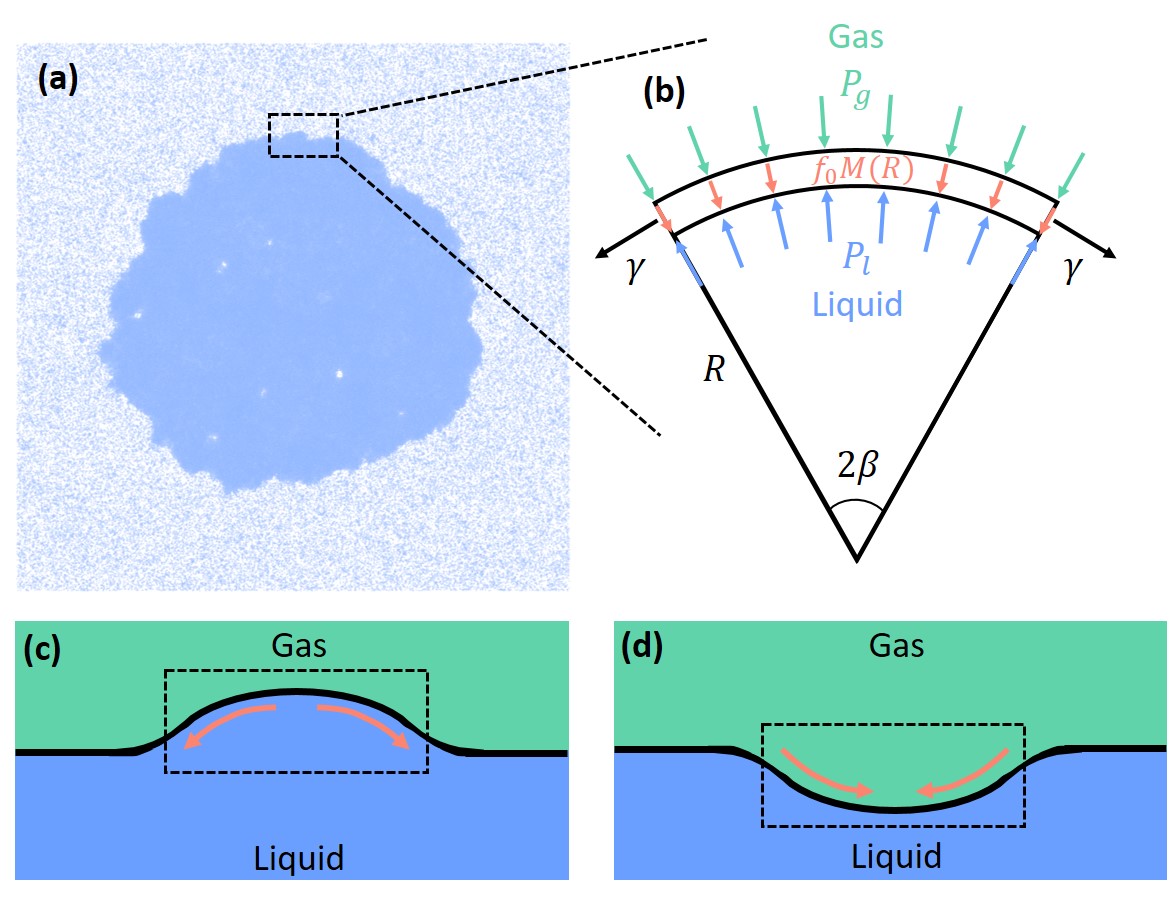}
	\caption{(a) Snapshot of a steady-state active droplet encompassed by the coexisting gas at $\text{Pe = 120}$. (b) Force analysis of a small arc segment of the droplet's surface, with a central angle $2\beta$. Here, the green, blue, black and pink arrows represent the gas pressure, liquid pressure, surface tension and polarization force $f_0 M(R)$, respectively. Sketch of a fluctuation-induced transient bulge (c) and indentation (d) on the macroscopic planar interface of the MIPS, where the arrows represent the possible particle flows or the corresponding tangential driving forces.}
	\label{Fig4}
\end{figure}

\textbf{B.2 Droplet}. Based on the mechanical definition of surface tension in the framework of intrinsic pressure, we have obtained positive $\gamma$ for the planar active interface. Next, we show that the significantly positive $\gamma$ is consistent with the mechanical stability of the active interface. To this end, we first simulate a steady-state droplet of radius $R$ encompassed by the coexisting gas within a square box, as illustrated in Fig.~\ref{Fig4}(a). The surface of the isotropic circular droplet is in mechanical equilibrium, and the forces on a differential segment of the surface are depicted in Fig.~\ref{Fig4}(b). Considering the emergence of the polarization force, the mechanical balance on the surface segment reads
\begin{equation}
	\label {eq11a}
	2\gamma\sin\beta + \int_{-\beta}^{\beta} [f_0 M(R) + P_g - P_l]R \cos\theta {\rm d} \theta = 0
\end{equation}
with $P_l$ referring to the pressure inside the droplet and $f_0M(R)$ to the polarization force per unit arc length acting on the droplet surface with a curvature radius $R$. Upon elimination of $\sin\beta$, Eq.~(\ref{eq11a}) is reduced to a generalized Young-Laplace equation,
\begin{equation}
	\label {eq11b}
	\frac{\gamma}{R} = P_l  - P_g - f_0 M(R).
\end{equation}

Equation~(\ref{eq11b}) provides an independent route to compute the surface tension. To compare with the aforementioned planar interface system, we here implement the simulations for a large system containing $N = 300000$ ABPs, such that the radius of the formed steady-state droplet is much larger than the interface's thickness. Considering the costly computations, without loss of the generality, we only focus on the case of $\text{Pe = 120}$ as a typical example. Based on the numerically measured quantities for the droplet system: $P_l \approx 129.6$, $P_g \approx 1.0$, $f_0 M(R) \approx 124.0$ and $R \approx 215 \sigma$,  Eq.~(\ref{eq11b}) yields $\gamma \approx 1957$. The mechanical equilibrium of the steady-state droplet surface thus requires a largely positive $\gamma$, although its magnitude is approximately twice as large as that of the planar interface case measured via Eq.~(\ref{eq6}). This difference in the magnitude reasonably arises from the curvature effects of the droplet and the significant fluctuations of the measured $M(R)$. It should be pointed out that in a recent work~\cite{Lauersdorf2021sm} Lauersdorf \textit{et al.} also employ a similar mechanical equilibrium route to compute $\gamma$, but their central formula contains an error and the uncertainty of their results is too large to determine the sign of $\gamma$.

\textbf{B.3 Fluctuating interface}. Besides maintaining the mechanical equilibrium of the steady-state droplet's surface, we further demonstrate that a significantly positive surface tension is also necessary for stabilizing a fluctuating active interface. Consider a horizontal planar interface of macroscopically phase-separated ABPs that has developed a bulge or indentation with a large curvature radius $R$ due to fluctuations, as sketched in Figs.~\ref{Fig4}(c) and (d). Unlike the steady-state circular droplet, this fluctuation-induced bulge rapidly relaxes to the stable flat interface. In addition, the nonuniform curvature of the transient bulge could be expected to give rise to tangential particle flows along its surface~\cite{Patch2018sm,Langford2024pre}, which may pratially suppress the interfacial fluctuations. The driving force for the tangential flow, that possibly originates from the nonvanishing tangential polarization, is proportional to the inhomogeneity of the interface curvature. Thus, the stability of the fluctuating interface requires the forces on the bulge to meet the following inequality,
\begin{equation}
	\label {eq8a}
	2\gamma\sin\beta + F_d + \int_{-\beta}^{\beta} [f_0 M(R) + P_g - P_l]R \cos\theta {\rm d} \theta > 0
\end{equation}
with $F_d$ denoting the contribution from the tangential driving force. It should be noted that here $P_g$ and $P_l$ refer to the pressures of the coexisting gas and liquid phases, respectively, separated by the planar interface. 

Given that the indentation has opposite local curvatures compared to the bulge, due to the symmetry the mechanical analysis of the transient indentation thus yields
\begin{equation}
	\label {eq8b}
	2\gamma\sin\beta - F_d + \int_{-\beta}^{\beta} [-f_0 M(-R) - P_g + P_l]R \cos\theta {\rm d} \theta > 0,
\end{equation}
where $f_0 M(-R)$ denotes the polarization force per unit arc length on the indentation, and the curvature dependence of $\gamma$ is ignored. By adding Eq.~(\ref{eq8a}) to Eq.~(\ref{eq8b}) and integrating $\theta$ out, in the leading-order approximation we have
\begin{equation}
	\label {eq8c}
	\gamma >  [f_0 M(-R) - f_0 M(R)]\frac{R}{2} \approx [f_0 M - f_0 M(R)]R.
\end{equation}
Note that $M(R)$ depends on the curvature radius of the surface and thus differs from the value of $M$ for the flat interface.

The inequality~(\ref{eq8c}) thus gives a lower limit of surface tension to stabilize the fluctuating interface. If the transient bulge and the stable droplet have the same curvature radius, their $M(R)$ would be approximately equal. Then, inserting the measured polarization of the flat interface, $f_0 M \approx 126.9$, and the above droplet's polarization $f_0 M(R)\approx 124$ (radius $R \approx 430$) into Eq.~(\ref{eq8c}), we have $\gamma > 1247$ for $\text{Pe} = 120$. This significantly positive $\gamma$ is in good agreement with the results obtained from Eqs.~(\ref{eq6}) and (\ref{eq11b}), and is needed to maintain a stable planar active interface of the MIPS.

\section*{Discussion}

From the analysis above, the large positive $\gamma$ can suppress the interfacial fluctuations and stabilize the active interfaces of the MIPS; while it seemingly is incompatible with the formation of bubbles inside the liquid phase (see Fig.~\ref{Fig2}(a)), a repeatedly reported phenomenon in the MIPS~\cite{Tjhung2018prx,Caporusso2020prl,Shi2020prl}. This apparent contradiction can be resolved by recognizing that both $\gamma$ and polarization force must be considered concurrently when examining the emergence of the bubbles in the MIPS, which is unlike the case for passive bubbles. As detailed in the Supplementary material~\cite{suppl}, the polarization force acting on the bubble surface can be strong enough to overwhelm the large positive $\gamma$, thereby enabling the bubble to grow to a significant size before it merges with other bubbles or leaves the liquid phase.

Finally, it is instructive to comment on the distinction between the expressions of mechanical surface tension of the MIPS derived from the intrinsic pressure and from the active pressure. Compared to the active pressure-based formula, which predicts a counterintuitive negative surface tension for the phase-separated ABPs, the formula we have derived within the intrinsic pressure framework yields a positive surface tension. This result aligns with physical intuition and well-established knowledge. Nevertheless, this does not exclude the usefulness of the active pressure framework or diminish its mechanical self-consistency~\cite{Solon2018njp}, provided that all relevant quantities are taken to conform to the framework of active pressure. In addition, there is a significant difference in the scope of application between the two frameworks. The surface tension based on active pressure framework is only limited to special active systems, i.e. spherical ABPs, where the active pressure is a state function. Otherwise, the coexisting phases possess unequal active pressures, such that the Kirkwood-Buff-type expression for the surface tension is inapplicable. In contrast, the framework of intrinsic pressure, which is inherently a state function, does not suffer from such limitations and is universally valid for all phase-separated active systems.

\section*{Conclusion}
On the basis of the intrinsic pressure, we present a new expression of surface tension of phase-separated active systems from its mechanical definition. According to the obtained formula, our simulation measurements show that the surface tension of the MIPS of the ABPs is significantly positive, in contrast to the previous studies~\cite{Bialke2015prl,Omar2020pre} that reported a largely negative or a near-zero surface tension. The present results align with the surface tension of the droplet determiend by the modified Young-Laplace equation and are also strongly supported by the mechanical stability analysis of the fluctuating active interface of the MIPS. The positive surface tension is naturally compatible with the stable MIPS, so our work provides a promising route to solve the puzzle of negative mechanical surface tension in the MIPS. Furthermore, our findings are not confined to the ABPs and can be straightforwardly applied to quantify surface tension for a wide range of other active systems, thereby paving the way for exploring diverse active interfacial phenomena. 

\section{Acknowledgment}
We thank Matthias Schmidt for helpful discussions. This work was supported by the National Natural Science Foundation of China (No. T2325027, 12274448).

%\bibliographystyle{apsrev}
%\bibliography{myref}

\end{document}